\documentclass[preprint,secnumarabic,amssymb, nobibnotes, aps, prd]{revtex4-1}

\usepackage{graphicx}

\begin{document}

\title{Noise Measurements of High-Speed, Light-Emitting GaN Resonant-Tunneling Diodes}%

\author{E. R. Brown}%
\email[Email: ]{elliott.brown@wright.edu}
\affiliation{Department of Physics, Wright State Univ., Dayton, OH  45435}
\author{W-D. Zhang}%
\email[Email: ]{wzzhang@fastmail.fm}
\affiliation{Department of Physics, Wright State Univ., Dayton, OH  45435}
\author{T. A. Growden}%
\affiliation{Department of Electrical and Computer Engineering, Ohio State Univ., Columbus, OH 43210}
\author{P. R. Berger}%
\affiliation{Department of Electrical and Computer Engineering, Ohio State Univ., Columbus, OH 43210}
\author{R. Droopad}%
\affiliation{Ingram School of Engineering, Texas State Univ., San Marcos, TX  78666}
\author{D. F. Storm}%
\affiliation{Electronics Science and Tech. Division, Code 6852, U.S. Naval Research Laboratory, Washington, DC  20375}
\author{D. J. Meyer}%
\affiliation{Electronics Science and Tech. Division, Code 6852, U.S. Naval Research Laboratory, Washington, DC  20375}

\date{\today}%

\begin{abstract}
We report here the first RF noise measurements on two designs of n-doped GaN/AlN double-barrier resonant tunneling diodes (RTDs), each having a room-temperature negative differential resistance (NDR) and also strong near-UV light emission.  The measurements are made with a standard, un-isolated RF receiver and calibration is made using a substitution-resistor/hot-cold radiometric technique which works in the positive differential resistance (PDR) region but not the NDR region.  A high-quality InGaAs/AlAs double-barrier RTD is used as a control sample and displays shot noise suppression down to $\Gamma\approx$0.5 in the PDR region, as expected.  The GaN/AlN RTDs display both shot-noise enhancement and suppression in the PDR regions, but no obvious sign of sudden shot-noise enhancement in the threshold bias region of light emission.  This supports the hypothesis that the holes required for light emission are created by electronic (Zener) interband tunneling, not impact ionization. Further the minimum shot-noise factor of $\Gamma\sim$ 0.34 suggests that the GaN/AlN RTDs are acting like triple-barrier devices.
\end{abstract}
\maketitle

\section{Introduction}
Thirty years ago, the double-barrier resonant-tunneling diode (RTD) was one of the most studied solid-state devices in the world.  Displaying a room-temperature negative differential resistance (NDR) and quantum-limited (tunneling) transport, the RTD offered a very high maximum frequency of oscillation $>$ 1.0 THz \cite{Brown1} and 2.0-ps-grade switching speed. \cite{Whitaker2}  This led to a number of interesting applications including a local oscillator in mm-wave-to-THz heterodyne receivers \cite{Blundell3} and fast switches in triggering circuits, \cite{Diamond4} to name a few.  Although the speed of the RTD in these applications was impressively fast, interest waned because of the longstanding problem of dc instability of RTDs in integrated circuits without isolation. In recent years, this problem has been mitigated by the monolithic integration of RTDs in special circuits, such as RTD oscillators coupled to resonant antennas. \cite{Asada5, Feiginov6} In the oscillator and other applications, a recurring question is the noise characteristics of RTD-based circuits because two-terminal devices, in general, do not provide input-output isolation.

In this paper, we report the first measurements of the noise behavior made on the latest version of RTDs – those fabricated from the wurtzite GaN/AlN materials system with only n-type doping on freestanding GaN substrates.  A recent university-led basic research effort has resulted in AlN double-barrier RTDs having stable and high-speed NDR,\cite{Growden7,Encomendero8} similar to that developed for GaAs/AlAs in the 1980s. \cite{Tsuchiya9} But it has taken many years to reach this point because the GaN/AlN heterostructures bring greater difficulty in the design, MBE growth, and fabrication.  This is largely because of the strong piezoelectric effects that occur at the heterointerfaces and spontaneous polarization that is inherent to the wurtzite structures. \cite{Ambacher10} The design rules for successful GaAs/AlAs RTDs do not work for GaN/AlN devices without special care in solving Poisson’s equation for the band bending in the presence of the huge polarization fields ($\sim$5$\times$10$^{6}$V/cm). In addition, the MBE growth (plasma assisted) is challenging, as is the dry etching of micron-scale devices. \cite{Storm11}

As an added benefit, these GaN/AlN RTDs have displayed pronounced light-emission effects not previously reported in GaAs/AlGaAs or similar unipolar-n-doped materials. \cite{Growden12} The strongest emission ($\sim$360 nm) has occurred in the near-UV just above the GaN bandgap, suggesting that a significant density of free holes is created somewhere in the device structure. Two possibilities for the hole generation are Zener (interband) tunneling and cross-gap impact ionization.  Both are mediated by the large internal electric fields in the devices and are difficult to discriminate from I-V measurements alone.  However, impact ionization always entails an enhancement of the current noise compared to normal (i.e., “full”) shot noise assuming the multiplication gain M is $>>$ 1. \cite{McIntyre13} This is routinely the case in impact ionization avalanche transit-time (IMPATT) diodes, \cite{Hines14} and avalanche photodiodes. \cite{Yuan15}A second reason for studying the noise behavior is that GaAs- and InP-based RTDs have long displayed interesting modification of the shot noise with suppression in the PDR regions and, when measurement is feasible, enhancement in the NDR region. So, an interesting question is whether or not GaN/AlN RTDs display a similar behavior.  And these are in addition to the widely-held belief that noise can reveal aspects of the device physics not seen in the time-averaged transport parameters, and subject to a famous quote. \cite{Branscomb16}
  
\section{Experiments and Results}

The measurement of noise in RTDs is a non-trivial exercise in circuit coupling, stability, and calibration.  Past measurements of GaAs- or InGaAs-based devices have been made with some type of circuit isolation.  Low RF measurements (typically below 1 MHz) have been made with the coupling to high-impedance voltage amplifiers, and on RTDs having very low peak- current-density ($J_{P}$) so that the differential conductance G was low enough to obtain dc-bias stability was achieved in the NDR region without oscillation. \cite{Li17, Iannaccone18, Kuznetsov19} Higher frequency measurements (typically 1 GHz and above) have been made with coupling to standard 50-ohm circuits but containing a microwave circulator \cite{Brown20} or isolator \cite{Przadka21}  between the RTD and the first amplifier. Both methods have displayed suppression of the shot noise in the PDR regions as represented by the generalized shot noise power spectral density, $S_{I}=2eI_{0}\Gamma$, where $e$ is the electron charge, $I_{0}$ is the average (dc) current, and $\Gamma$ is the shot-noise factor ( $<$1.0 for suppression). Shot noise enhancement ($\Gamma>$1.0) has also been observed in the NDR region but only in the low-$J_{P}$ devices \cite{Yuan15} or in devices oscillating with high-enough spectral purity to estimate $\Gamma$ from the frequency noise.\cite{Li17} 

\begin{figure*}[tb]
\includegraphics[scale=0.50]{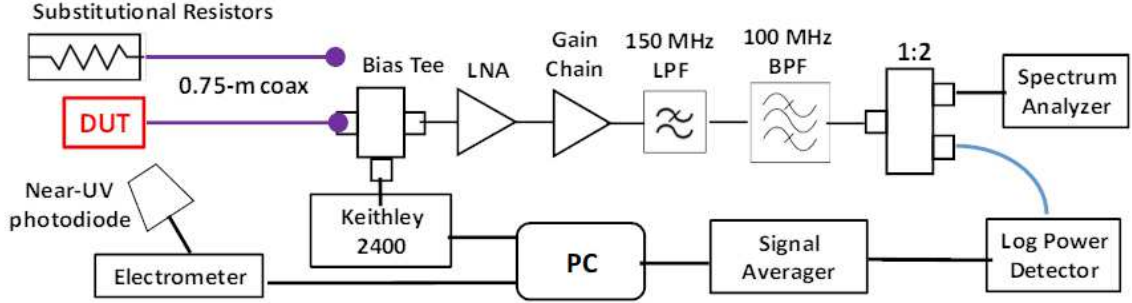}
\caption{\label{fig:1}Block diagram of experimental set-up for measuring RF noise and near-UV electroluminescence simultaneously.  The DUT is the GaN/AlN RTD device under test.}
\end{figure*} 

The present method is based on the un-isolated RF receiver circuit shown in Fig. \ref{fig:1}.  The unpackaged RTD is located on the platen of a probe station and contacted with 40-GHz ground-signal-ground probes.  Bias voltage is applied through a wideband bias tee and the RF port of the bias tee is connected to a first-stage LNA through a 0.75-m-long coaxial cable.  The LNA is followed by more amplifiers and then two filters: (1) a high-Q bandpass filter centered at 100 MHz, and (2) a low-pass filter to block RF power above 150 MHz.  The 100 MHz center frequency was chosen to be well above the 1/f noise of the RTDs and other components in Fig. \ref{fig:1}, but well below the frequency where they become reactive through their internal capacitance.  The output of the filters is fed into a logarithmic power detector having an operating range between $\sim$ -60 and -30 dBm.  The equivalent noise bandwidth (ENBW) of the amplifier-chain plus filters was 3.7 MHz.  The output of the power detector was connected to a signal averaging DC voltmeter, and the RTD under test was biased with a Keithley 2400 set in voltage-source mode.  Meanwhile, the light emission from the RTD was monitored with a sensitive Si photodiode coupled optically to the device with a brass light pipe.  The photodiode current was measured with an electrometer (accurate down to $\sim$ 1 pA).

Calibration of the receiver was carried out by a substitution-resistor, hot-cold (SRHC) method.  A set of metal-film resistors spanning the range of 10 Ω to 4.7 kΩ were connected directly to coaxial lines in shielded boxes.  Each was connected directly to the bias tee of Fig. \ref{fig:1} and the output power was measured both at room temperature ($\sim$295 K) and at 77 K, the latter measurement made possible by immersing the resistor box in liquid nitrogen.  High-quality metal film resistors are known to have low inductance and small change in resistance (typically 2\%) between $\sim$295 and 77 K.  By the Y-factor method, \cite{Keysigntnote22} we derived the gain and noise temperature of the receiver in the narrow 100-MHz-centered passband as a function of source resistance. Then with the RTD device in place, the receiver contribution to the total noise power subtracted out, leaving just the noise contribution from the RTD.  Finally, the RTD noise was analyzed with an equivalent circuit model for the RTD modeled as a differential conductance G in shunt with the shot noise current generator and in series with a contact resistance.  This model ignores device capacitance, which is generally negligible at 100 MHz in the RTDs tested here.  The thermal-noise contribution of the contact resistance was found to be similarly negligible.  This calibration procedure was not applicable to the NDR region because of the lack of any negative substitution resistor.

\begin{figure*}[tb]
\includegraphics[scale=0.35]{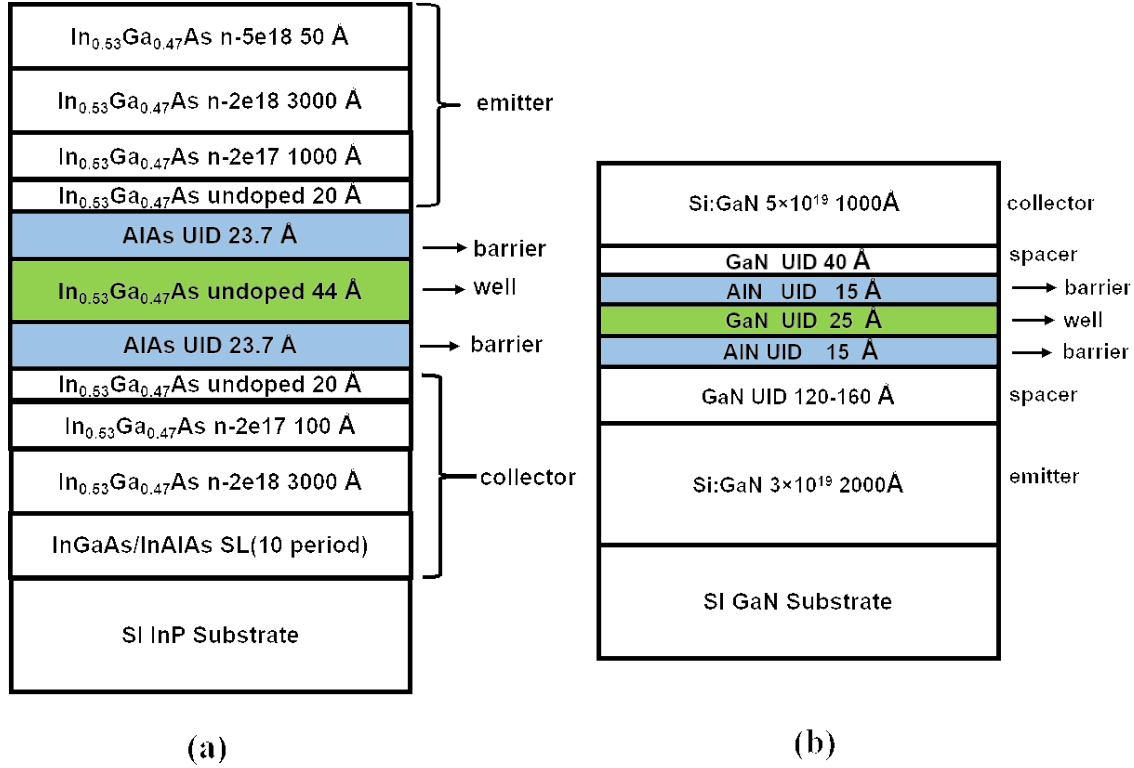}
\caption{\label{fig:s}Structures of RTDs: (a)InGaAs/AlAs, and (b)GaN/AlN. The unit for doping densities is cm$^{-3}$.}
\end{figure*} 

\begin{figure*}[tb]
\includegraphics[scale=0.55]{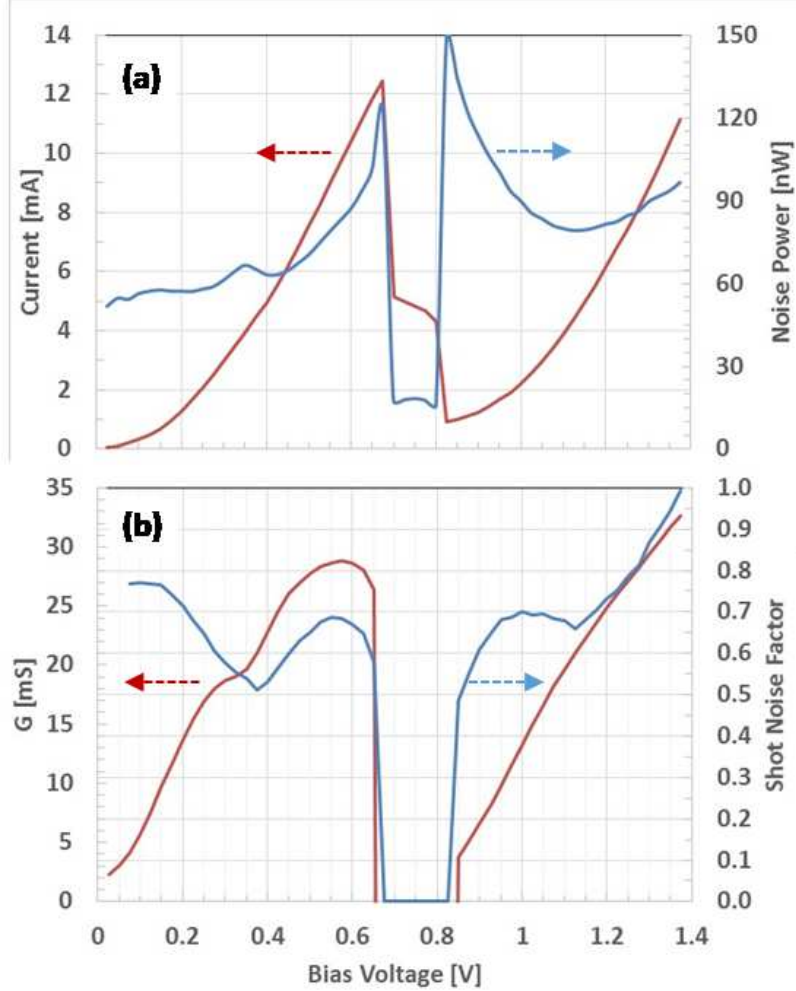}
\caption{\label{fig:3}Experimental results for high quality InGaAs/AlAs RTD at room temperature and plotted vs bias voltage: (a) Electrical current (left axis), and noise power (right axis).  (b) Differential conductance (left axis), and shot-noise factor (right axis).}
\end{figure*} 

As a proof-of-concept, the calibrated receiver was first demonstrated on a control sample consisting of an RTD of predictable behavior: a high quality In$_{0.53}$Ga$_{0.47}$As/AlAs RTD [Fig. \ref{fig:s}(a)] having moderate $J_{P}$ ($\approx$ 20 kA/cm$^{2}$) previously-measured for its fast switching speed. \cite{Growden23} The upward-going I-V curve and RTD-specific noise power are plotted in Fig. \ref{fig:3}(a), and the differential conductance and shot-noise factor in Fig. \ref{fig:3}(b).  The I-V curve displays a peak-to-valley current ratio of PVCR$\approx$12 and a chair-like behavior in the NDR region characteristic of a circuit oscillation.  The noise-power vs. bias voltage is monotonic except for a dropout of power in the NDR region and peak just below and above it.  The oscillation in the NDR region is occurring at a frequency well above the 100-MHz passband (as observed with the spectrum analyzer in Fig. \ref{fig:1}), but it is strong enough to saturate the amplifier chain and decrease the noise power in the passband accordingly. The peak in power just below the NDR is associated with the rapidly increasing current (and shot noise), whereas the peak above it is attributed mostly to the low differential conductance around the valley point. Everywhere else in the PDR bias regions the shot noise is readily analyzed, and the shot-noise factor plotted in Fig. \ref{fig:3}(b).   As expected from all previous studies of GaAs- and InGaAs-based double-barrier RTDs, the shot noise is suppressed ($\Gamma<$ 1.0) in the PDR regions with a minimum $\Gamma\approx$ 0.5.   At a bias just below 0.4 and above 1.1 V are not attributed to any intrinsic noise effect but rather to points where the receiver gain reached a maximum associated with the impedance match between the RTD and LNA.

\begin{figure*}[tb]
\includegraphics[scale=0.45]{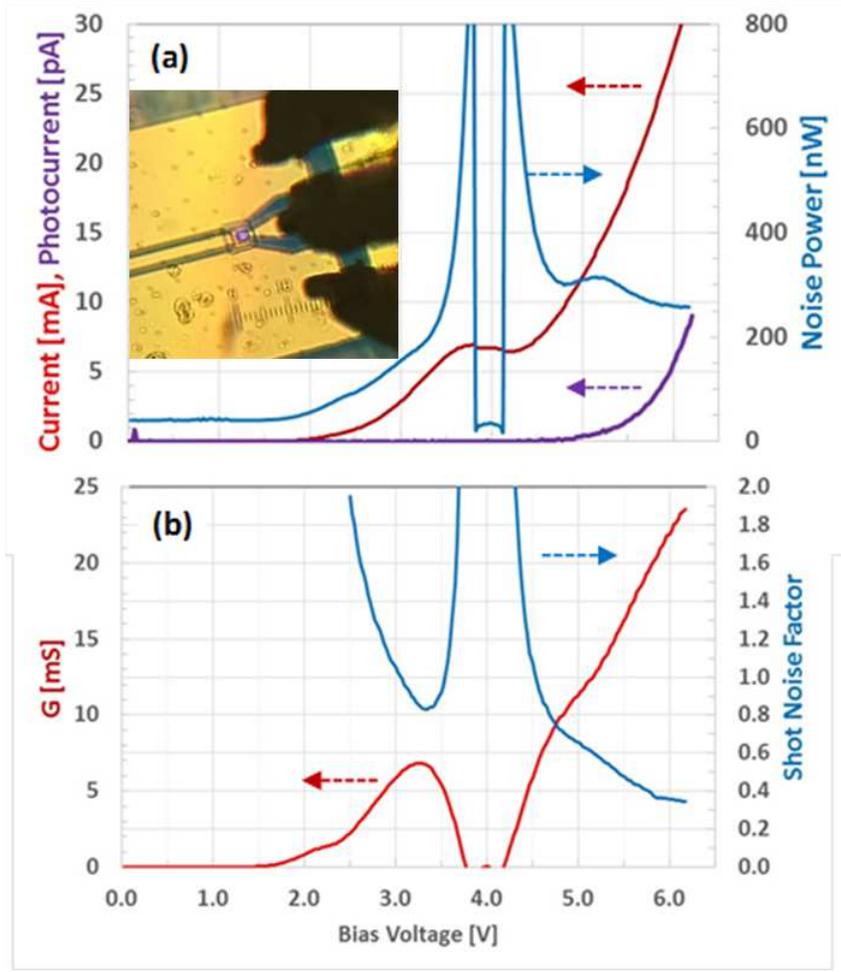}
\caption{\label{fig:4}Experimental results for moderate current density GaN/AlN RTD at room temperature and plotted vs bias voltage: (a) Electrical current and photocurrent (left axis), and noise power (right axis); (b) Differential conductance (left axis) and shot noise factor (right axis).  The inset in (a) shows the RTD device-under-test emitting violet light (long-wavelength tail of spectrum centered in the near-UV).}
\end{figure*} 

We then connected a moderate-$J_{P}$ ($\approx$ 2.7$\times$10$^{3}$ A/cm$^{2}$) GaN/AlN double-barrier RTD [Fig. \ref{fig:s}(b)]\cite{Growden7} through a GSG probe, and monitored the light emission [inset of Fig. \ref{fig:4} (a)].  The resulting I-V, L-V and noise power curves are plotted in Fig. \ref{fig:4}(a) where the bias voltage was kept below 6.2 V to avoid device burn-out.  The NDR region occurs between $\sim$3.8 and 4.2 V and displays a PVCR of $\approx$ 1.08 along with the chair-like structure indicative of oscillations. The threshold for strong light emission occurs at $\sim$4.8 V.  The noise power is similar to the InGaAs RTD displaying pronounced peaks adjacent to the NDR region and also the oscillation-induced drop-out in the NDR region.  But well away from the NDR region, the behavior is more complicated as evidenced by the obvious peak in noise power around 5.2 V.  This corresponds to an inflection point in the G-vs-V curve in Fig. \ref{fig:4}(b), and after the noise analysis occurs in a region where the shot noise factor is decreasing to $<$ 0.4.  There is no sign of a sudden increase in the shot noise near the $\sim$ 4.8 V threshold of light emission, which suggests that the mobile holes required for the cross-gap are created by a normal-shot-noise mechanism, such as Zener tunneling, not impact ionization.  Interestingly, at bias voltages well below the NDR region (e.g. $\sim$3 V), the shot noise is enhanced ($\Gamma>$1.0) for reasons that are not yet understood.

\begin{figure*}[tb]
\includegraphics[scale=0.65]{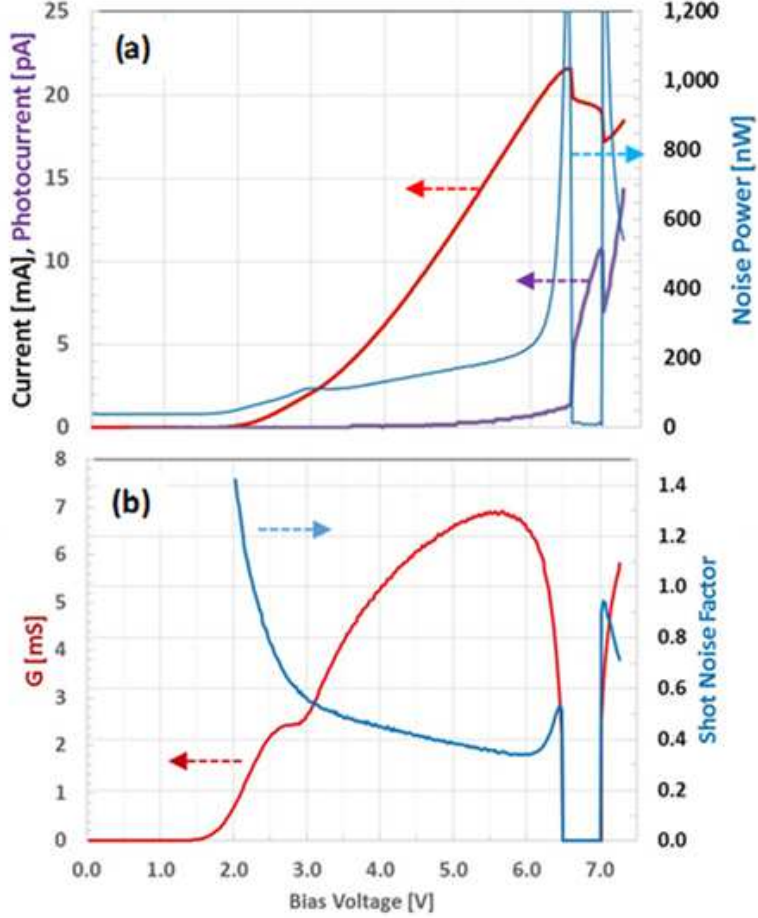}
\caption{\label{fig:5} Experimental results for high-current-density GaN/AlN RTD at room temperature and plotted vs bias voltage: (a) Electrical current and photocurrent (left axis), and noise power (right axis), (b) Differential conductance (left axis), and shot noise factor (right axis).}
\end{figure*} 

Finally, we connected a high-$J_{P}$ ($\approx$2.0$\times$10$^{5}$ A/cm$^{2}$) GaN/AlN double-barrier RTD in the same manner as above. \cite{Growden24} The resulting I-V, L-V and noise power curves are plotted in Fig. \ref{fig:5}(a) where the bias voltage was kept below 7.3 V to avoid device burn-out.  The NDR regions occurs between $\sim$6.5 and 7.0 V and displays a PVCR of $\approx$ 1.26 along with the chair-like structure indicative of oscillations. The threshold for strong light emission occurs at $\sim$6.5 V – in the midst of the NDR region. The noise power is similar qualitatively to the moderate-$J_{P}$ GaN/AlN RTD but is larger in magnitude, consistent with the higher current and shot noise of this device. At low bias voltages, the shot noise is enhanced too (for example $\sim$2.1V). The peaks adjacent to the NDR region are also less prominent, consistent with a closer impedance match to the LNA, but the noise in the NDR region drops out once again. At a given bias voltage the light emission is weaker than in the moderate-$J_{P}$ device, and displays a more abrupt threshold.  But this might be an artifact of its coincidence with the NDR region or due to the structure difference of the two RTDs. \cite{Growden12, Brown20} Another distinction of the high-$J_{P}$ device is the behavior below the current peak.  The high-$J_{P}$ device shows its greatest degree of shot-noise suppression [Fig. \ref{fig:5}(b)] at $\approx$6.0 V with $\Gamma\approx$0.34.  Figure \ref{fig:5}(a) also shows an obvious local peak in the noise power around 3.0 V, and \ref{fig:5} (b) a local chair at the same bias.  As described in a separate manuscript, this is attributable to electron tunneling through the lowest quasibound state E$_{1}$ of the GaN RTD, meaning that the primary peak at 6.4 V is likely electron tunneling through E$_{2}$. \cite{Pouyet25} 	

\section{Summary}
In conclusion, we have performed the first-known measurements of the noise behavior of GaN/AlN RTDs and related it to the electrical (I-V) and light-emission (L-V) characteristics.  We observe no definitive sign of noise enhancement at the light-emission threshold, consistent with the required hole generation being by Zener tunneling, not impact ionization.  On the other hand, the shot noise does display substantial suppression in the PDR regions of the device – above the NDR region in the low J$_{P}$ device, and below it in the high J$_{P}$ device.  The shot-noise factor of $\sim$0.34 is below the double-barrier (Fano) limit of 0.5, but close to the triple-barrier limit of 0.33. \cite{Newaz26} GaN/AlN RTDs are known to have deep “pre-wells” ($\sim$0.5-0.7 eV) on the emitter side of the structure and a high “pre-barrier” ($\sim$0.1-0.2 eV) separating the AlN barriers from the quasi-neutral region on the emitter side.  If true, this would suggest that the GaN/AlN devices studied here are acting like triple-barrier devices – a result made possible by noise measurements and modeling.


This work was sponsored by a MURI Program (DATE) under Dr. Paul Maki and the U.S. Office of Naval Research. The first author acknowledges Dr. Patrick Fay of Notre Dame University for helpful comments about noise measurements.

\end{document}